\documentclass[a4paper,11pt]{article}
\usepackage{jcappub} 

\usepackage{amsmath}
\usepackage{amssymb}
\usepackage{graphicx}
\usepackage{color}
\usepackage{subfig}
\usepackage{bm}
\usepackage{gensymb}
\usepackage{bigints}
\usepackage[usenames,dvipsnames,svgnames,table]{xcolor}
\usepackage{aas_macros}
\usepackage{orcidlink}
\usepackage{physics}
\usepackage[sort&compress]{natbib}

\arxivnumber{2308.16604} 
\title{\boldmath Gravitational Lensing in Modified Gravity: A case study for Fast Radio Bursts}

\author[1]{Surajit Kalita\orcidlink{0000-0002-3818-6037},\note{Corresponding author}}

\affiliation{High Energy Physics, Cosmology \& Astrophysics Theory (HEPCAT) Group, \\ Department of Mathematics \& Applied Mathematics, University of Cape Town, \\ Cape Town 7700, South Africa}

\emailAdd{surajit.kalita@uct.ac.za}

\author{Shruti Bhatporia\orcidlink{0000-0003-2821-4927}}
\emailAdd{bhtshr001@myuct.ac.za}
\author{and Amanda Weltman\orcidlink{0000-0002-5974-4114}}
\emailAdd{amanda.weltman@uct.ac.za}

\abstract{Over the last few decades, a plethora of modifications to general relativity have been proposed to solve a host of cosmological and astrophysical problems. Many modified gravity models are now ruled out with further astrophysical observations; some theories are still viable, with, at best, bounds on their parameters set by observations to date. More recently, observations of Fast Radio Bursts have proven to be remarkably powerful tools to constrain cosmology and fundamental physics. In this work, we consider a generic modified gravity theory and consider the implications for gravitational lensing with Fast Radio Bursts. We use a set of Fast Radio Burst observations to constrain the fraction of dark matter made up of primordial black holes in such a theory. We further show that modified gravity adds a screening effect on gravitational lensing similar to the case when there is plasma in the path of the light ray acting as a scattering screen.}

\begin{document}
\maketitle
\flushbottom

\section{Introduction}

Over the last century, Einstein's theory of General Relativity (GR) has proven its success as a theory of gravity; explaining astronomical observations ranging from the perihelion precession of Mercury to the prediction and subsequent observation of gravitational waves (GWs) from compact object mergers. GR has been well-tested through observations of the solar system and nearby (low redshift) objects. However, on cosmological scales, the theory seems to fall short, requiring new ingredients to explain the observation of the accelerated expansion of the universe today, and the inflationary phase in the early universe~\cite{1999ApJ...517..565P}. The theory also predicts a small-scale singularity and is not a renormalizable theory~\cite{1974AIHPA..20...69T}. As a result, it is still important to test gravity theories in the strong-field limit. Moreover, GR is not adequate to explain massive white dwarfs and neutron stars, which were respectively inferred from the observations of peculiar overluminous type Ia supernovae~\cite{2006Natur.443..308H,2010ApJ...713.1073S} and LIGO/Virgo GW merger events~\cite{2020ApJ...896L..44A,2020ApJ...904...39H}.

To solve these astrophysical and cosmological problems, there is a proliferation of different theories with modifications to GR. Buchdahl first proposed nonlinear terms, such as $R^2$, $R^{\alpha\beta}R_{\alpha\beta}$, and $R^{\alpha\beta\mu\nu}R_{\alpha\beta\mu\nu}$ with $R$, $R^{\alpha\beta}$, $R^{\alpha\beta\mu\nu}$ respectively being the Ricci scalar, Ricci tensor, Riemann tensor, in the Lagrangian to examine their effects on the modified Friedmann equation for different matter compositions~\cite{1970MNRAS.150....1B}. A detailed analysis on the renormalization of higher-derivative gravity was given in~\cite{1977PhRvD..16..953S}. Later, Starobinsky used the $R^2$ correction term to explain the inflationary epoch of the universe~\cite{1980PhLB...91...99S}, which has recently been confirmed by the Planck data~\cite{2022EPJC...82..506B}. In more recent years, different modified gravity models have been explored not only in cosmological regimes but also in astrophysical contexts, such as to study compact object physics, including explaining massive white dwarfs~\cite{2018JCAP...09..007K,2022PhRvD.106l4010A} and neutron stars~\cite{2020PhR...876....1O}, the possibility of detecting additional modes of GWs~\cite{2011PhRvD..83j4022B,2021ApJ...909...65K}, etc. Note that in the case of modified gravity, apart from tensor modes, there can be additional massive scalar and vector modes, whose masses depend on the background. Since these massive modes are exponentially damped, current GW detectors cannot detect them~\cite{2021JCAP...03..014L}, and hence it is not straightforward to verify modified gravity theories from GW observations. However, using LIGO/Virgo merger events, constraints were put on the speed of GWs ($c_\text{gw}$); for example, with the observed time delay between the GW merger event GW\,170817 and the gamma-ray burst event GW\,170817A, it was found that $-3\times 10^{-15} < c_\text{gw}/c-1 < 7\times10^{-16}$ where $c$ is the speed of light~\cite{2017ApJ...848L..13A}. It was later shown that the Horndeski theory satisfies the bounds provided the mass scale $m$ satisfies $2\times10^{-35} \lesssim m \ll 10^{15}\rm\, GeV$~\cite{2018EPJC...78..738G}. Note that in this work, we only consider the massive modes with different spins. For the massless case, the situation could be different as discussed in~\cite{2018PhRvD..97l4023O,2021CQGra..38s5003G} for constraining the Einstein-{\ae}ther theory.

On the other hand, Fast Radio Burst (FRB) observations since 2007~\cite{2007Sci...318..777L}, have opened up a new window in astronomy. FRBs are bright, short-lived transient events observed in the radio spectrum within approximately the 100\,MHz to 8\,GHz frequency range~\cite{2019PhR...821....1P}. So far, there are over 700 FRBs observed with the majority of them observed with the purpose-built Canadian Hydrogen Intensity Mapping Experiment (CHIME)\footnote{\url{https://chime-experiment.ca/en}} radio telescope. The expected rate of FRBs across the entire sky is estimated to be approximately $100-1000$ per day~\citep{2016MNRAS.460L..30C}. With the remarkable exception of FRB\,200428, which is verified to have originated from a Galactic magnetar SGR\,$1935+2154$~\citep{2020PASP..132c4202B,2020Natur.587...59B,2020Natur.587...54C}, their relatively large dispersion measures (DMs) imply that the majority of these FRBs have extragalactic origins.

Ample different models have been proposed to explain the physical mechanism driving FRBs~\cite{2019PhR...821....1P}, most of which incorporate white dwarfs, neutron stars, or black holes. These theories can broadly be classified into two categories. First, the merger of these compact objects where coherent radio emission is produced at the time of the merger~\cite{2014A&A...562A.137F,2013PASJ...65L..12T,2013ApJ...776L..39K,2020IJAA...10...28L}. The second category broadly involves mechanisms that do not require any merger event and some physical activity in the compact object can cause the emission of radio waves. Examples of these mechanisms include the curvature radiation mechanism~\citep{2017MNRAS.468.2726K,2019MNRAS.483L..93L}, starquake mechanisms such as the crustal activity of a magnetar~\citep{2018ApJ...852..140W}, synchrotron maser emission from relativistic, magnetized shocks~\citep{2014MNRAS.442L...9L}, giant flares in soft gamma repeaters~\citep{2014ApJ...797...70K}, Gertsenshtein-Zel'dovich effect~\cite{2023MNRAS.520.3742K}, etc. Detailed reviews of progenitor theories can be found in~\cite{2019PhR...821....1P} and~\cite{2020Natur.587...45Z}. Most FRBs seem to be non-repeating (one-off bursts) while some do repeat albeit with no known periodicity. As a result, progenitor theories that predict repeating FRBs seem more promising because they can explain the apparently one-off bursts, indicating that they may repeat after a long period or we have not yet seen their repetitions, possibly due to selection biases in our observations. Of course, it is possible, even likely, that there are simply multiple classes of FRBs and this work is agnostic on that front.

Due to their distinct features, such as very short pulse widths, large DMs, and the fact that they probe the intergalactic medium on cosmological scales, FRBs have been used to study different astrophysical and cosmological questions; a non-exhaustive list follows. Using FRB\,150418, it was earlier found that the constraint on the photon mass $m_\gamma<1.8\times 10^{-14}\rm\,eV\,c^{-2}$~\cite{2016PhLB..757..548B} and this bound is now stronger over time considering many other FRBs~\cite{2021PhLB..82036596W,2023MNRAS.520.1324L}. Moreover, using 16 localized and 60 unlocalized FRBs, the Hubble constant $H_0$ was estimated to be $H_0 = 73^{+12}_{-8}\rm\,km\,s^{-1}\,Mpc^{-1}$~\cite{2022MNRAS.516.4862J}. Note that the error bars are large because of the relatively few FRBs used. Furthermore, using the concept of gravitational lensing in FRBs, Mu{\~n}oz et al. showed for the first time that the constraint on the fraction of dark matter made up of primordial black holes is $f_\mathrm{PBH}<0.08$ assuming some FRBs are lensed by the black hole of mass $M>20\,M_\odot$~\cite{2016PhRvL.117i1301M}. Later these bounds were improved considering FRB microstructures~\cite{2020ApJ...900..122S} and extended mass functions~\cite{2020PhRvD.102b3016L}. With this technique, Liao et al. visualized this constraint using 110 real FRB data with null detection of lensed FRBs~\cite{2020ApJ...896L..11L}. More recently, using 172 bursts from 114 CHIME FRBs, it was shown that the $f_\mathrm{PBH}$ bound can change significantly if there is plasma in the path of the light ray which also acts as a decoherence or scattering screen~\cite{2022PhRvD.106d3017L}. In this paper, we investigate the effect of modified gravity on the $f_\mathrm{PBH}$ bound using the non-detection of lensing of 636 CHIME FRBs. We show that modified gravity behaves similarly to plasma lensing, which acts as a screen in the path of the light ray. They both complicate the task of accurately deriving the best constraints. In Section~\ref{Sec 2}, we derive the time delay between the lensed images in modified gravity, which we then use in Section~\ref{Sec 3} to obtain the $f_\mathrm{PBH}$ bound using CHIME FRBs. Finally, Section~\ref{Sec 4} discusses these results and provides some parting thoughts as conclusions.

\section{Time delay in modified gravity}\label{Sec 2}

Assuming the metric signature $(+,-,-,-)$, in the spherical polar coordinates $(t, r, \theta, \phi)$, let us consider the following spherically symmetric line element
\begin{align}
    \dd{s}^2 = B(r)c^2\dd{t}^2 - A(r)\dd{r}^2 - r^2\dd{\theta}^2 - r^2\sin^2\theta\dd{\phi}^2.
\end{align}
Hence, the time required for the light ray to reach the distance of closest approach $r_0$ from an arbitrary distance $r$, or vice-versa, is given by~\cite{1972gcpa.book.....W}
\begin{align}\label{Eq: time delay}
    t\left(r,r_0\right) = \frac{1}{c}\bigintsss_{r_0}^r \sqrt{\frac{A(r)/B(r)}{1-\frac{B(r)}{B(r_0)}\left(\frac{r_0}{r}\right)^2}}\dd{r}.
\end{align}
In GR, for the Schwarzschild metric outside of a black hole of mass $M_\mathrm{L}$, expanding up to $\mathcal{O}(M_\mathrm{L}/r)$, the above expression can be written as follows~\cite{1972gcpa.book.....W}:
\begin{align}
    t\left(r,r_0\right) &= \frac{1}{c}\sqrt{r^2-r_0^2} + \frac{2GM_\mathrm{L}}{c^3}\ln(\frac{r+\sqrt{r^2-r_0^2}}{r_0}) + \frac{GM_\mathrm{L}}{c^3}\sqrt{\frac{r-r_0}{r+r_0}},
\end{align}
where $G$ is Newton's gravitational constant.

\begin{figure}[htpb]
    \centering
    \includegraphics[scale=0.3]{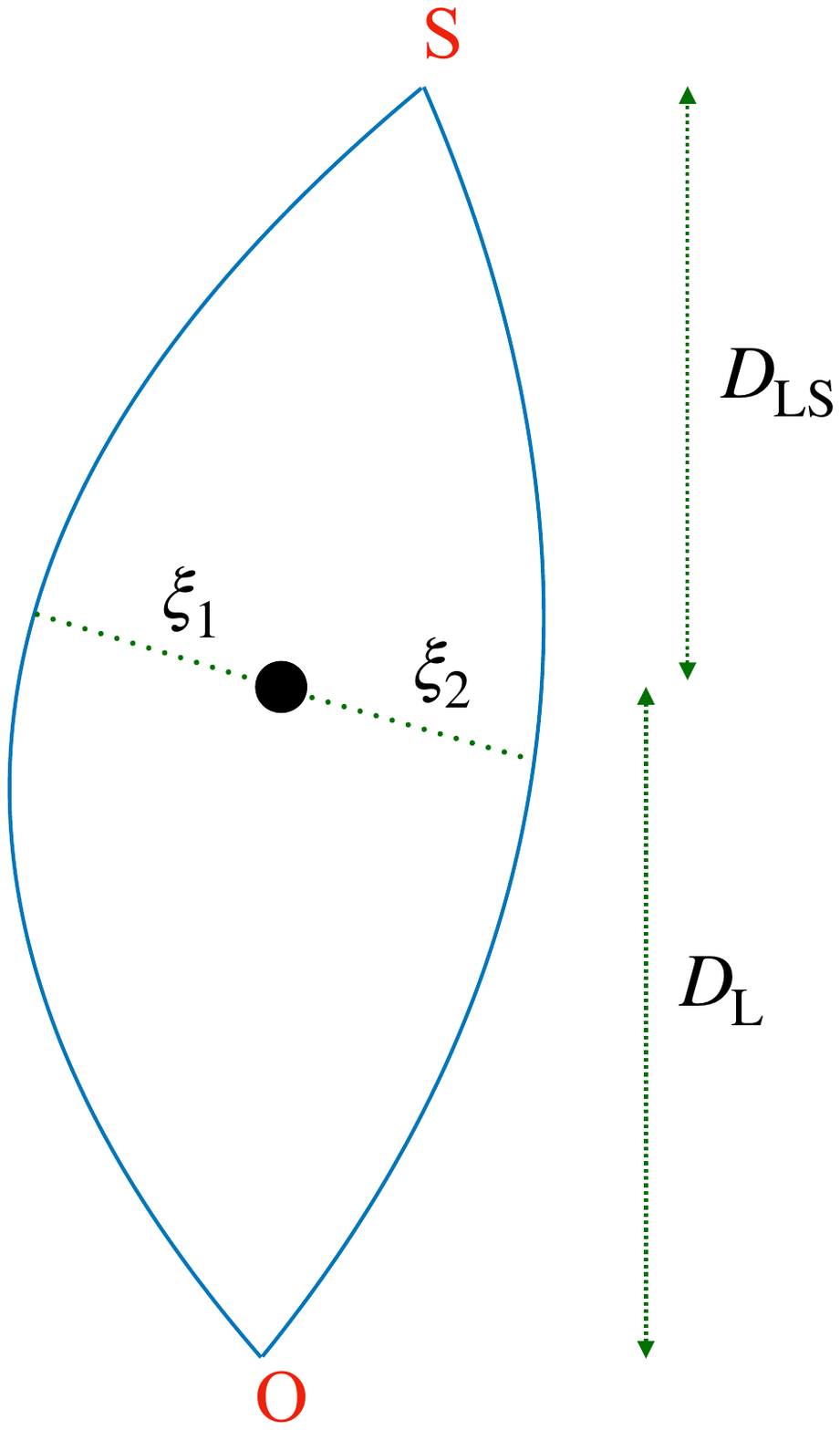}
    \caption{Schematic diagram of deflection of a light ray in the presence of a compact object.}
    \label{Fig: lensing}
\end{figure}
Now consider that a light ray travels from a source at S to an observer at O, passing by a compact object as shown in figure~\ref{Fig: lensing}. Due to the curvature of spacetime, the compact object behaves like a lens, and the light ray is deflected forming two images of the source at S for the observer at O. It is worth noting that here we are only considering the case of strong lensing where multiple images can be formed for a single source. Weak lensing and microlensing, where only a single deformed image is formed, are not considered here as we are interested in measuring the differences in delays between the two lensed images. Let us assume that the closest distances of these light rays from the compact object are $\xi_1$ and $\xi_2$, and the angular diameter distances of the compact object from O and S are respectively $D_\mathrm{L}$ and $D_\mathrm{LS}$. Therefore, defining the Einstein radius $\theta_\mathrm{E}$ as
\begin{align}
    \theta_\mathrm{E} &= \sqrt{\frac{4GM_\mathrm{L}}{c^2}\frac{D_\mathrm{LS}}{D_\mathrm{L}\left(D_\mathrm{L}+D_\mathrm{LS}\right)}},
\end{align}
we define the following dimensionless quantities
\begin{align}
    y &= \frac{r}{\theta_\mathrm{E}\left(D_\mathrm{L}+D_\mathrm{LS}\right)}\\
    \text{and} \quad \zeta_{1,2} &= \frac{\xi_{1,2}}{D_\mathrm{L}\theta_\mathrm{E}},
\end{align}
such that the positions of the lensed images are given by
\begin{align}
    \zeta_{1,2} = \frac{1}{2}\left(y \pm \sqrt{y^2+4}\right).
\end{align}
If the redshift of the lens is $z_\mathrm{L}$, the differential time delay between the two images is given by
\begin{align}
    \Delta t = \left(1+z_\mathrm{L}\right) \left[t\left(D_\mathrm{LS},\xi_1\right) + t\left(D_\mathrm{L},\xi_1\right) - t\left(D_\mathrm{LS},\xi_2\right) - t\left(D_\mathrm{L},\xi_2\right) \right].
\end{align}
Assuming $\xi_1,\xi_2\ll D_\mathrm{L},D_\mathrm{LS}$, for the Schwarzschild metric, the above expression can be approximated as~\cite{1985A&A...148..369S}
\begin{align}\label{Eq: diff time delay}
    \Delta t \approx \frac{4G M_\mathrm{L}\left(1+z_\mathrm{L}\right)}{c^3} \left[\frac{y}{2}\sqrt{y^2+4} + \ln(\frac{\sqrt{y^2+4} + y}{\sqrt{y^2+4} - y}) \right].
\end{align}

Let us now consider the following generic modified spherically symmetric metric with modified gravity parameter $\Psi$ capturing the effect of modified gravity 
\begin{align}\label{Eq: MG metric}
    \dd{s}^2 &= \left(1-\frac{2GM_\mathrm{L}}{c^2r}+\Psi r^2\right)c^2\dd{t}^2 - \frac{1}{1-\frac{2GM_\mathrm{L}}{c^2r}+\Psi r^2}\dd{r}^2 -r^2\dd{\theta}^2-r^2\sin^2\theta\dd{\phi}^2.
\end{align}
It is also known as the Schwarzschild--de Sitter (SdS) metric. In GR, the Schwarzschild metric is a unique solution of the Einstein field equation according to Birkhoff's theorem. However, if one incorporates the cosmological constant $\Lambda$ in the field equation, the solution resembles the aforementioned metric of equation~\eqref{Eq: MG metric}. It is worth noting that this solution no longer readily reduces to the flat Minkowski metric in the limit $r\to\infty$. Moreover, this metric is also a solution for different modified gravity theories. For example, in $f(R)$ gravity, it was shown that the above metric is a solution when $R$ remains constant i.e. $R=R_0$ provided $R_0f'(R_0) - 2f(R_0)=0$~\cite{2006PhRvD..74f4022M}. In such a case, we have $\Psi = R_0/12$. Note that by replacing $f(R)$ by $f(X)-f'(X)(R-X)$ with $X$ being a dynamical scalar field and defining $\phi=-f'(X)$, one can show that the action of $f(R)$ gravity resembles the same for scalar-tensor theory provided $f''(X)\neq 0$. Moreover, one can define a potential $V(\phi)$ such that it follows $\phi^2 V = \phi X(\phi)-f(X(\phi))$~\cite{2018tegp.book.....W}. This potential provides the scalar field with a non-zero effective mass in the presence of matter via the `chameleon mechanism'~\cite{2004PhRvL..93q1104K,2004PhRvD..69d4026K}. The mass of this scalar field depends on the curvature and thus density of the system. Therefore, $f(R)$ gravity turns out to be a subset of chameleon gravity. Furthermore, the SdS metric is also a solution in $f(T)$ gravity with $T$ being the scalar torsion~\cite{2013EPJC...73.2272A}.

Now, substituting the SdS metric in equation~\eqref{Eq: time delay} and expanding up to $\mathcal{O}(M/r)$, we obtain
\begin{align}
    t\left(r,r_0\right) &= \frac{\tan^{-1}\left(\sqrt{\Psi} \sqrt{\frac{r^2-r_0^2}{1+\Psi r_0^2}}\right)}{c\sqrt{\Psi}} + \frac{2GM_\mathrm{L}}{c^3}\tanh^{-1}\left[r \sqrt{\frac{1+\Psi r_0^2}{r^2-r_0^2}}\right] \nonumber \\ &+ \frac{GM_\mathrm{L}}{c^3}\frac{(1-\Psi r r_0)}{\sqrt{1+\Psi r_0^2}\left(1+\Psi r^2\right)} \sqrt{\frac{r-r_0}{r+r_0}}.
\end{align}
Therefore, the differential time delay between the two images is given by
\begin{align}
    \Delta t &\approx \frac{1+z_\mathrm{L}}{c\sqrt{\Psi}}\tan^{-1} \left[\frac{2\sqrt{\Psi}\left(D_\mathrm{L}+D_\mathrm{LS}\right)\left(\xi_1^2-\xi_2^2\right)\left(2-\Psi\xi_1^2\right)\left(2-\Psi\xi_2^2\right)}{D_\mathrm{L}D_\mathrm{LS}\left(2-\Psi\xi_1^2\right)^2\left(2-A\xi_2^2\right)^2 - 4\Psi\left(\xi_1^2-\xi_2^2\right)^2}\right] \nonumber \\ &+ \frac{4GM_\mathrm{L}\left(1+z_\mathrm{L}\right)}{c^3} \ln(\frac{\sqrt{y^2+4} + y}{\sqrt{y^2+4} - y}).
\end{align}
In the limit $\Psi\xi_1^2, \Psi\xi_2^2 \ll 1$, it reduces to
\begin{align}
    \Delta t &=  \frac{1+z_\mathrm{L}}{c\sqrt{\Psi}}\tan^{-1} \left[\frac{\sqrt{\Psi}\left(D_\mathrm{L}+D_\mathrm{LS}\right)\left(\xi_1^2-\xi_2^2\right)}{2D_\mathrm{L}D_\mathrm{LS}}\right] + \frac{4GM_\mathrm{L}\left(1+z_\mathrm{L}\right)}{c^3}\ln(\frac{\sqrt{y^2+4} + y}{\sqrt{y^2+4} - y})\\
    &= \left(1+z_\mathrm{L}\right) \left[\frac{1}{c\sqrt{\Psi}}\tan^{-1} \left(\sqrt{\Psi}\frac{4GM}{c^2} \frac{y}{2}\sqrt{y^2+4}\right) + \frac{4GM_\mathrm{L}}{c^3} \ln(\frac{\sqrt{y^2+4} + y}{\sqrt{y^2+4} - y})\right].
\end{align}
As mentioned above, we are only interested in the case of strong lensing and this limit ensures that the light ray passes close enough to the black hole which can produce multiple images. It is evident that this expression reduces to equation~\eqref{Eq: diff time delay} as $\Psi\to0$. Moreover, the magnification ratio of these two images is given by
\begin{align}
    \mu = \left(\frac{y+\sqrt{y^2+4}}{y-\sqrt{y^2+4}}\right)^2.
\end{align}
\begin{figure}[htpb]
    \centering
    \includegraphics[scale=0.3]{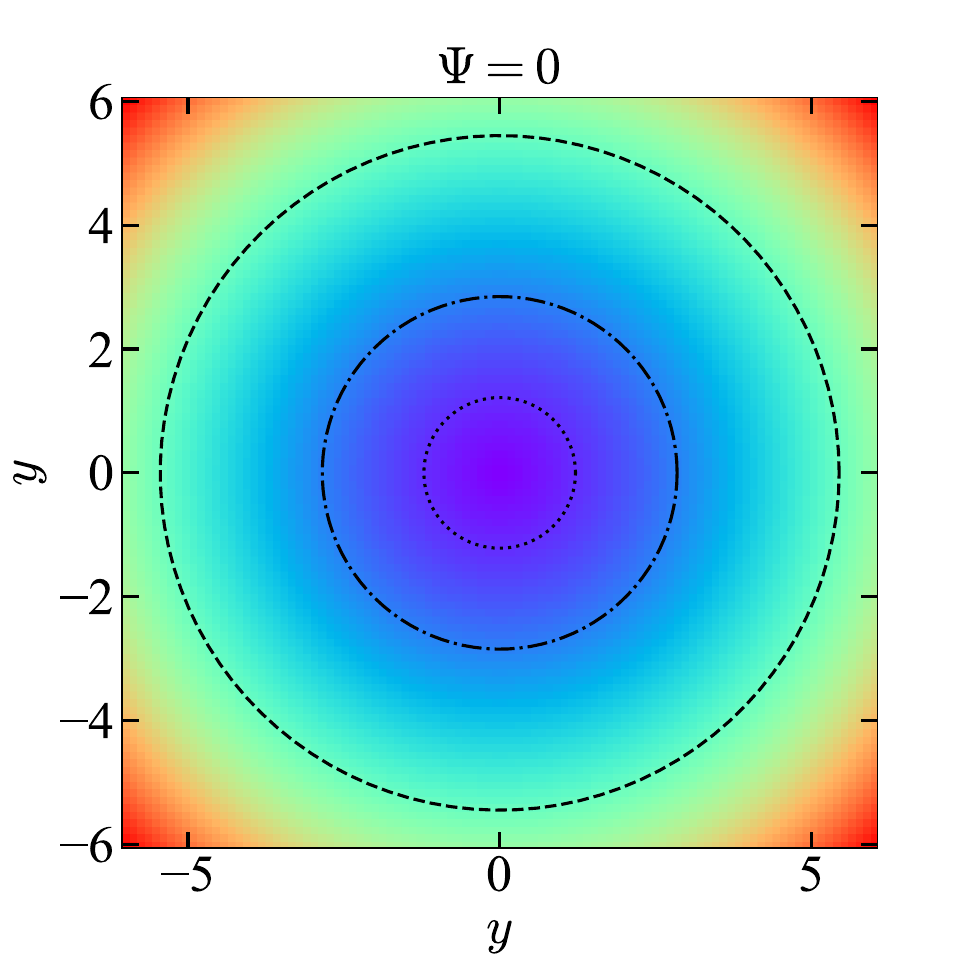}
    \includegraphics[scale=0.3]{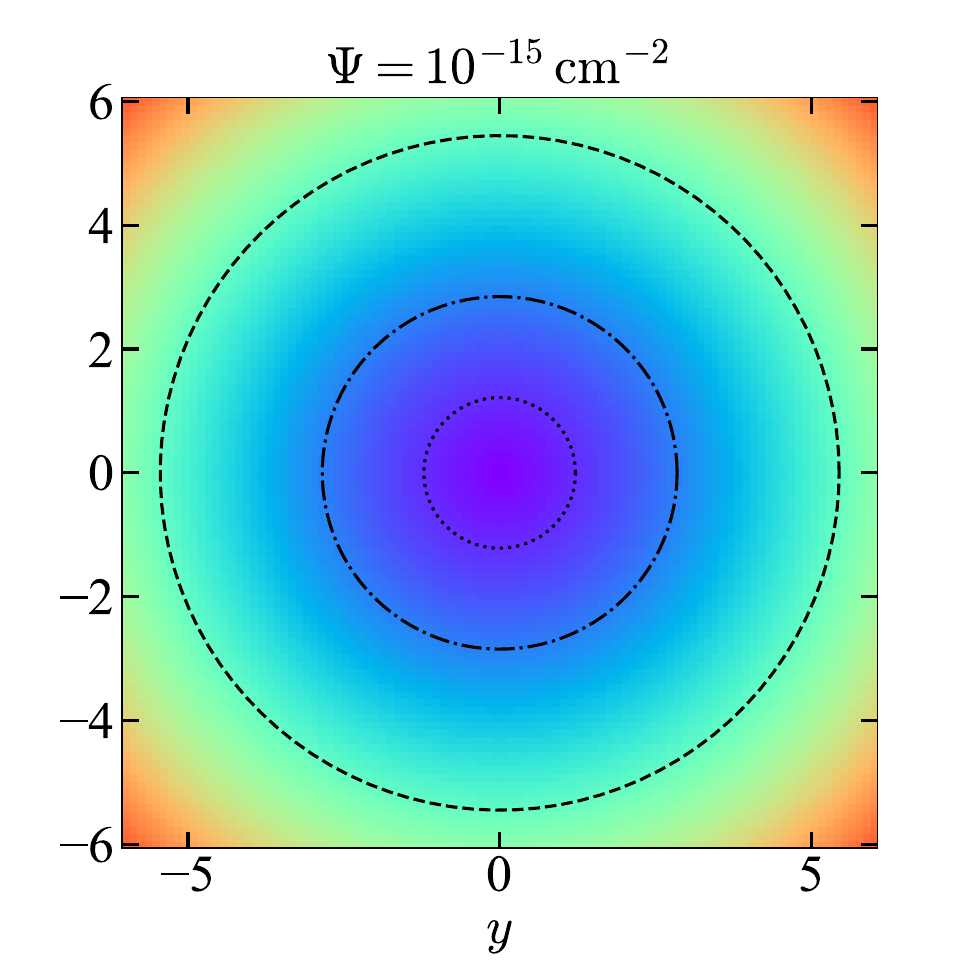}
    \includegraphics[scale=0.3]{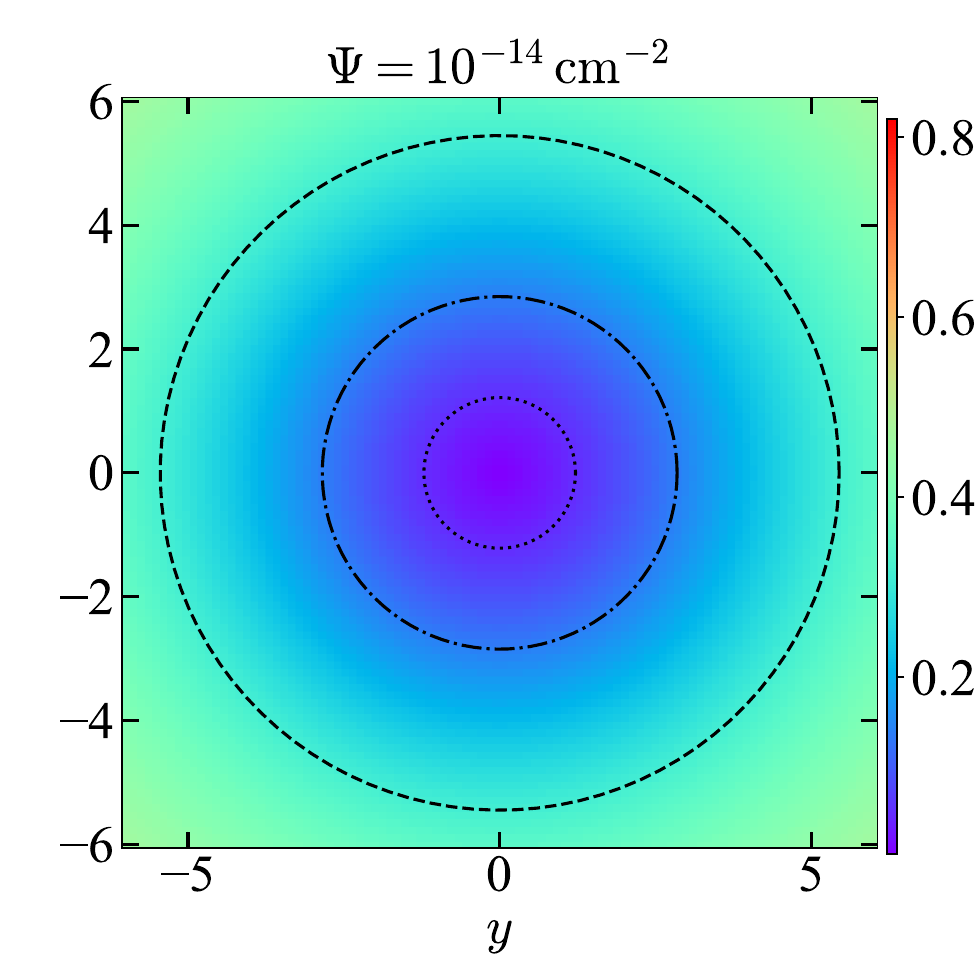}
    \caption{Differential delay $\Delta t$ for $1\,M_\odot$ lens mass as $\Psi$ increases. The color bar shows $\Delta t$ in the units of ms. The dotted, dash-dotted, and dashed circles depict the radii of the delay surfaces corresponding to $\mu=10$, $100$, and $1000$, respectively.}
    \label{fig: lensing delay}
\end{figure}

Figure~\ref{fig: lensing delay} shows different time delay surfaces as $\Psi$ changes assuming $z_\mathrm{L}=0$. The dotted, dash-dotted, and dashed circles indicate the amount of time delay $\Delta t$ between the two lensed images for different magnifications. It is evident that as $\Psi$ increases, $\Delta t$ at the same radius decreases. In other words, while obtaining the same magnification of the lensed images, the time difference between the two rays to reach the observer is reduced. This is because as $\Psi$ increases, the effective mass of the lens decreases; thus producing less gravitational curvature around it, and the light rays are effectively deflected less. In some sense, the presence of modified gravity somewhat screens the compact object.

\section{Effect of modified gravity in lensing of fast radio bursts}\label{Sec 3}

In this article, we consider 636 FRBs detected by CHIME\footnote{as of April 2023}. We are limited by the dimmest image which should also be bright enough to be detected, which sets an upper limit in $\mu$. Denoting $\text{S/N}$ to be the signal-to-noise ratio of the primary peak of the burst, it was inferred that the bounds on $\mu$ were $1<\mu<1/3\times \text{S/N}$~\cite{2020ApJ...900..122S}. Assuming the redshift of the source $z_\mathrm{S}$, its optical depth is given by~\cite{2016PhRvL.117i1301M}
\begin{align}\label{Eq: optical depth}
    \tau(M_\mathrm{L},z_\mathrm{S}) = \int_0^{z_\mathrm{S}} \dd{\chi(z_\mathrm{L})} \left(1+z_\mathrm{L}\right)^2 n_\mathrm{L} \sigma(M_\mathrm{L},z_\mathrm{L}),
\end{align}
where $\chi(z)$ is the comoving distance at a redshift $z$, $n_\mathrm{L}$ is the average comoving number density of the lens, and $\sigma$ is the lensing cross section for a point mass lens, given by the annulus between the minimum and maximum impact parameters, $y_\mathrm{min}$ and $y_\mathrm{max}$, respectively. Mathematically, it is represented by
\begin{align}\label{Eq: sigma}
    \sigma(M_\mathrm{L},z_\mathrm{L}) = \frac{4\pi G M_\mathrm{L} D_\mathrm{L} D_\mathrm{LS}}{c^2 D_\mathrm{S}} \left[y_\mathrm{max}^2(\mu) - y_\mathrm{min}^2(M_\mathrm{L},z_\mathrm{L})\right].
\end{align}
Note that as we are interested in constraints on primordial mass black holes making up Dark Matter, we are not restricted to solar mass black holes. Rather we want to consider primordial mass black holes, and hence we denote $M_\mathrm{L}$ as the mass of the primordial mass black hole. 
The comoving distance $\chi(z_\mathrm{L})$ is given by
\begin{equation}\label{Eq: chi}
    \chi(z_\mathrm{L}) = c\int_0^{z_\mathrm{L}} \frac{\dd{z}}{H(z)},
\end{equation}
where $H(z)$ is the Hubble function at a redshift $z$, given by $H(z) = H_0 \sqrt{\Omega_\mathrm{m} \left(1+z\right)^3 + \Omega_\Lambda}$ with $\Omega_\mathrm{m}$ and $\Omega_\Lambda$ respectively being the present matter and vacuum density fraction. Moreover, assuming a monochromatic mass function with a peak at mass $M_\mathrm{L}$, we define the following lens mass function $\dv*{n_\mathrm{L}}{M}$ as~\cite{2022PhRvD.106d3017L}
\begin{align}\label{Eq: lens mass function}
    \dv{n_\mathrm{L}}{M} = \frac{\rho_\text{crit}}{M_\mathrm{L}}\, f_\mathrm{PBH}\, \Omega_\mathrm{c}\, \delta(M-M_\mathrm{L}),
\end{align}
such that
\begin{align}\label{Eq: lens mass function 2}
    \int \dv{n_\mathrm{L}}{M} M \dd{M} = \rho_\text{crit}\, f_\mathrm{PBH}\, \Omega_\mathrm{c}.
\end{align}
Here $f_\mathrm{PBH}$ is the fraction of dark matter made up of primordial mass black holes, which act as the lens, $\Omega_\mathrm{c}$ is the current cold dark matter density, and $\rho_\text{crit}=3H_0^2/8\pi G$ is the critical density of the universe. Now, substituting equations~\eqref{Eq: sigma}--\eqref{Eq: lens mass function 2} in equation~\eqref{Eq: optical depth}, we obtain
\begin{align}\label{Eq: tau}
    \tau(M_\mathrm{L},z_\mathrm{S}) = &\frac{3}{2}f_\mathrm{PBH} \Omega_\mathrm{c}\int_0^{z_\mathrm{S}} \dd{z_\mathrm{L}} \frac{H_0^2}{cH(z_\mathrm{L})} \frac{D_\mathrm{L} D_\mathrm{LS}}{D_\mathrm{S}} \left(1+z_\mathrm{L}\right)^2 \left[y_\mathrm{max}^2(\mu) - y_\mathrm{min}^2(M_\mathrm{L},z_\mathrm{L})\right].
\end{align}

\begin{figure}[htpb]
    \centering
    \includegraphics[scale=0.7]{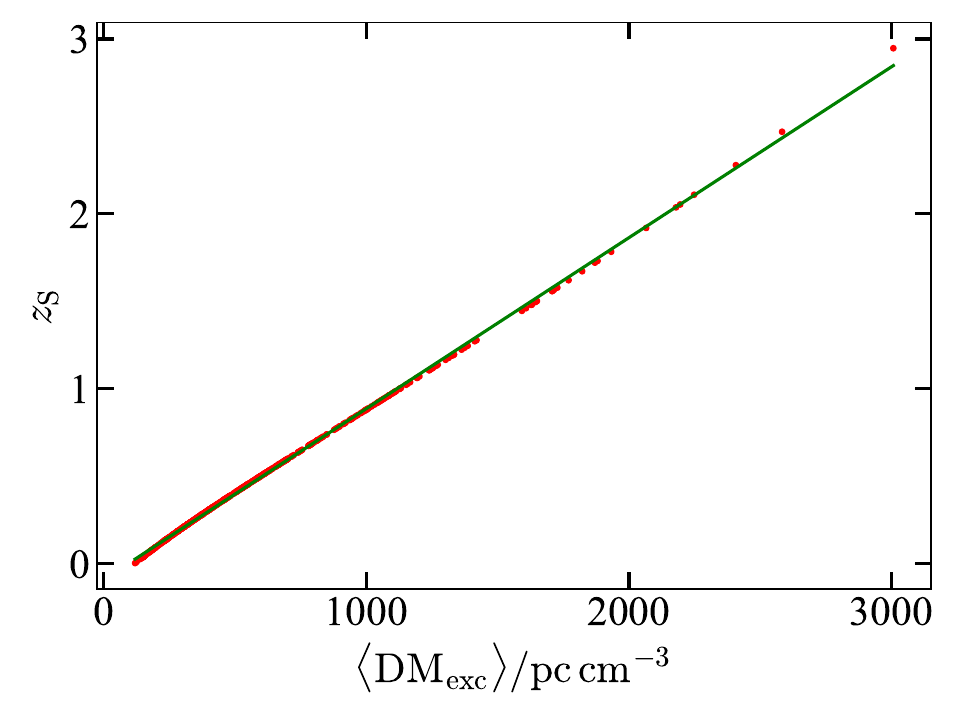}
    \caption{Source redshift as a function of average excess DM for the reported CHIME/FRB data sample shown in red dots. The green solid line is the best-fitted linear approximation with $z_\mathrm{S} \approx {\langle \text{DM}_\text{exc}(z_\mathrm{S})\rangle}/1023\rm\,pc\,cm^{-3}$.}
    \label{fig: Redshift estimates}
\end{figure}

The determination of redshifts of FRBs using their measured DM values is important for this work, thus we explain it here. The current CHIME/FRB dataset~\cite{2021ApJS..257...59C, Fonseca_2020, The_CHIME_FRB_Collaboration_2023} reports the DM as well as the excess DM  of FRBs, which we use to estimate the redshifts of these FRBs. The excess DM receives contributions from the host galaxy, the IGM, and foreground galaxies if any, and is found by subtracting the DM contribution of the Milky Way using the NE2001 model~\cite{2002astro.ph..7156C}. Observation/selection biases and propagation effects introduce noise into the CHIME/FRB data set, therefore we introduce cutoffs suggested by the CHIME/FRB team for our sample data set.  The average excess DM ($\langle \text{DM}_\text{exc}(z_\mathrm{S})\rangle$) measured for the bursts has two major contributions: from the inter-galactic medium (IGM) and from the host galaxy. Hence, based on~\cite{Walters_2018}, $\langle \text{DM}_\text{exc}(z_\mathrm{S})\rangle$ is given by
\begin{align}\label{Eq: DM excess}
    \langle \text{DM}_\text{exc}(z_\mathrm{S})\rangle &= \langle \text{DM}_\text{IGM}(z_\mathrm{S})\rangle + \langle \text{DM}_\text{Host}(z_\mathrm{S})\rangle ,
\end{align}
where following~\cite{2022PhRvD.106d3017L}, the median DM contribution from host galaxies is assumed to be $117\rm\, pc\, cm^{-3}$ at the source of the burst, such that
\begin{align}
\langle \text{DM}_\text{Host}(z_\mathrm{S})\rangle = \frac{117\rm\, pc\, cm^{-3}}{1 + z_\mathrm{S}}.
\end{align}
Here, $\langle \text{DM}_\text{IGM}(z)\rangle$ is the average DM contribution from the IGM along the line of sight. It is primarily a function of the electron density along the line of sight:

\begin{align}
    \langle \text{DM}_\text{IGM}(z_\mathrm{S})\rangle &= \frac{3cH_0\Omega_\mathrm{b}}{8\pi G m_\mathrm{p}} \int_{0}^{z_\mathrm{S}} \frac{f_\text{IGM}(z)\chi(z)(1 + z)}{\sqrt{\Omega_\mathrm{m} \left(1+z\right)^3 + \Omega_\Lambda}} \dd{z},
\end{align}
where $\Omega_\mathrm{b}$ is the baryonic matter density, $m_\mathrm{p}$ is the proton mass, $f_\text{IGM}$ is the baryon mass fraction in the IGM, and $\chi(z)$ the ionization fraction along the line of sight, given by
\begin{align}
    \chi(z) = Y_\mathrm{H}\chi_{\mathrm{e},\mathrm{H}}(z) + \frac{1}{2}Y_\mathrm{p}\chi_{\mathrm{e},\mathrm{He}}(z), 
\end{align}
with $\chi_{\mathrm{e},\mathrm{H}}$ and $\chi_{\mathrm{e},\mathrm{He}}$ respectively being the ionization fractions of the intergalactic hydrogen and helium, and $Y_\mathrm{H} = 3/4$, $Y_\mathrm{p} = 1/4$ their respective mass fractions. The factor $\sqrt{\Omega_\mathrm{m} \left(1+z\right)^3 + \Omega_\Lambda}$ encodes information about $\Lambda$ cold dark matter~($\Lambda$CDM) cosmology\cite{2020MNRAS.498.3927H}. We use the nine-year Wilkinson Microwave Anisotropy Probe~(WMAP9) for cosmological parameter values, assuming flat $\Lambda$CDM cosmology~\cite{2013ApJS..208...19H}.  We obtain $z_\mathrm{S}$ using equation~\eqref{Eq: DM excess}, shown in figure~\ref{fig: Redshift estimates}. Except for the sources with high DM, the DM--$z_\mathrm{S}$ relation is approximately linear, and for our analysis, $z_\mathrm{S} \approx {\langle \text{DM}_\text{exc}(z_\mathrm{S})\rangle}/1023\rm\,pc\,cm^{-3}$ holds, also depicted in figure~\ref{fig: Redshift estimates}. We include high DM FRBs considering their offsets from the linear relation described in~\cite{2021ApJ...922...42R}. CHIME/FRB team reported DM error bars based on \texttt{fit\_b} algorithm~\cite{2021ApJS..257...59C}, although future Very Long Baseline Interferometry~(VLBI) observations will provide a better estimation for DM error values. For this analysis, we use average DM values for the redshift estimates. We further consider a lower cutoff of ${\langle \text{DM}_\text{exc}(z_\mathrm{S})\rangle}$ at $117\rm\, pc\, cm^{-3}$, however the results are robust to the choice of cutoff. Moreover, these $ z_\mathrm {S}$ values, derived using equation~\eqref{Eq: DM excess}, are based on the measured average DMs; the true redshifts for some of these sources can be different from these estimated numbers depending on variations in the line of sight IGM, foreground galaxies if any or different host galaxy contributions.

Once we obtain $\tau$ for one FRB using equation~\eqref{Eq: tau}, we can compute the integrated optical depth $\Bar{\tau}$ considering all of the FRBs. If we know the redshift distribution function $N(z_\mathrm{L})$, it is given by
\begin{align}
    \bar{\tau}(M_\mathrm{L}) = \int \tau(M_\mathrm{L},z_\mathrm{L}) N(z_\mathrm{L}) \dd{z_\mathrm{L}}.
\end{align}
Previous works~\cite{2016PhRvL.117i1301M,2020PhRvD.102b3016L,2022PhRvD.106d3017L} used to consider different redshift distribution functions such as constant-density redshift distribution~\cite{2016MNRAS.461..984O} or star-formation redshift distribution~\cite{2016MNRAS.458..708C}. The constant-density redshift distribution is based on the assumption that FRBs have a constant comoving number density while the latter assumes that the FRBs follow the star-formation history. However, a recent result using the CHIME/FRB dataset suggests that $N(z_\mathrm{L})$ does not follow these distributions~\cite{2022ApJ...924L..14Z}. Moreover, we no longer require such an assumption because we now have data on a significant number of FRBs. Hence, we rather simply use the following discretized definition for the integrated optical depth
\begin{align}
    \bar{\tau} = \frac{1}{\mathcal{N}_\mathrm{FRB}}\sum_{i=1}^{\mathcal{N}_\mathrm{FRB}} \tau (M_\mathrm{L},z_{\mathrm{S},i}),
\end{align}
where $\mathcal{N}_\mathrm{FRB}$ is the number of FRBs in the data sample to be considered.

Since the lensed FRBs $\mathcal{N}_\mathrm{lensed, FRB}$ are expected to be quite small in number in comparison to the total number of FRBs, one can safely use Poisson statistics to obtain the following relation~\cite{2020PhRvD.102b3016L}
\begin{align}
    \mathcal{N}_\mathrm{lensed, FRB} = \left(1-e^{-\bar{\tau}}\right) \mathcal{N}_\mathrm{FRB}.
\end{align}
Owing to the fact that no lensed FRB has been confirmed so far, using the above relation, we obtain
\begin{align}
    f_\mathrm{PBH}<\frac{1}{\tau_1}\ln(\frac{\mathcal{N}_\mathrm{FRB}}{\mathcal{N}_\mathrm{FRB}-1}),
\end{align}
where $\bar{\tau} = f_\mathrm{PBH}\tau_1$. In the limit $\mathcal{N}_\mathrm{FRB}\gg1$, this relation reduces to $f_\mathrm{PBH}<1/(\tau_1\mathcal{N}_\mathrm{FRB})$.

\begin{figure}[htpb]
    \centering
    \includegraphics[scale=0.7]{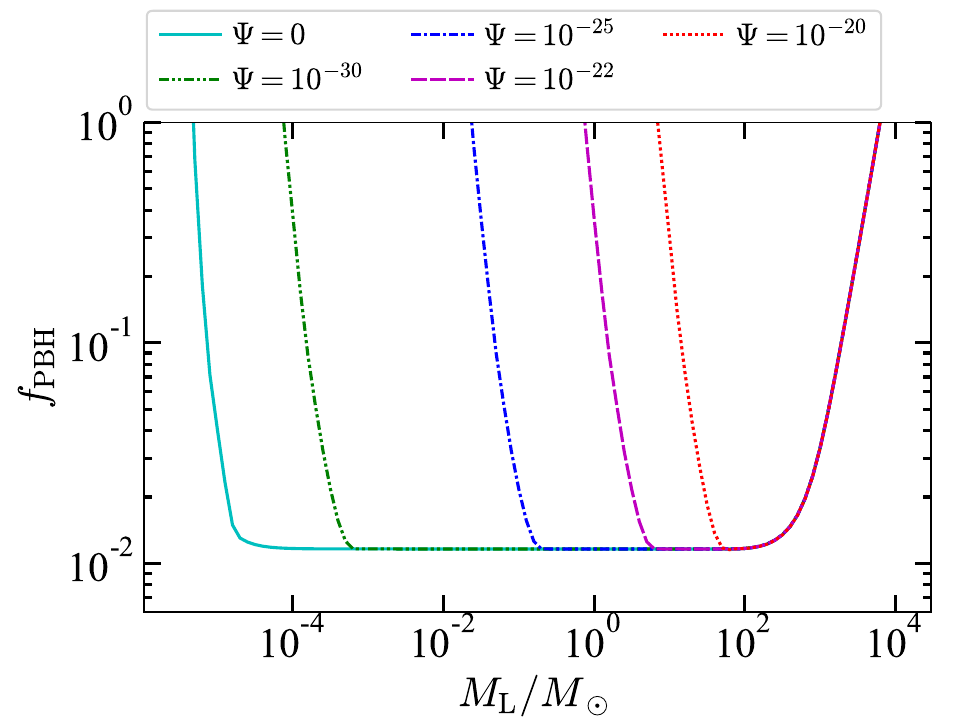}
    \caption{Bounds on the fraction of dark matter made up of primordial black holes for different values of modified gravity parameter $\Psi$ in the unit of cm$^{-2}$.}
    \label{Fig: f bound}
\end{figure}

Figure~\ref{Fig: f bound} shows the bound on $f_\mathrm{PBH}$ using the CHIME data for different modified gravity parameters. It is evident that as the value of $\Psi$ increases, the bound is changed considerably. The left cut-off of each curve is determined from the minimum $\Delta t$ of the telescope, which is approximately $10^{-9}\rm\,s$ for the CHIME telescope. The cut-off on the right side of each curve is determined from the maximum magnification ratio $\mu$, which essentially indicates if the dimmer image is detectable along with the brighter one. In this work, we choose $\mu=1/3\times \text{S/N}$ following~\cite{2020ApJ...900..122S}. For a certain lens mass, we obtain that $f_\mathrm{PBH}$ increases as $\Psi$ increases. In other words, the bound on $f_\mathrm{PBH}$ gets weaker with the increase in the modified gravity parameter. The modified gravity parameter $\Psi$ is constrained for various modified gravity models. The ongoing New Horizons mission in the outskirts of our solar system is providing extended radio-tracking data, which gives $\abs{\Psi}\lesssim10^{-49}\rm\,cm^{-2}$~\cite{2016PDU....13..111I}. It is important to note that these findings are highly dependent on the specific experiment and can vary with each observation. Different bounds on $\Psi$ were discussed in~\cite{2006PhLB..634..465K, 2006PhRvD..73f3004S} and found a much-relaxed bound with up to $\abs{\Psi}\lesssim 2\times 10^{-24}\rm\,cm^{-2}$. This clearly shows the chameleon effect~\cite{2004PhRvL..93q1104K,2004PhRvD..69d4026K} is present and the effective values could be very different at different length scales. In this work, we show how the bound on the fraction of dark matter made up of primordial black holes changes with changing $\Psi$. The exact values are not very important, rather the effect itself is what we hope to exemplify.

\section{Discussion and conclusion}\label{Sec 4}
The use of gravitational lensing to study modified gravity is not new, but it has not been studied in great detail from its timing perspective. Some properties of gravitational lensing was studied under $f(R,T)$~\cite{2016IJMPD..2550020A} and Horndeski theories~\cite{2023arXiv230408141B}. Moreover, a ray-tracing approach was utilized to construct weak lensing maps from the simulated light cones, enabling the examination of weak lensing effects under different gravity conditions~\cite{2018MNRAS.481.2813G}. More recently, using the SgrA* and M87* black hole shadow observations, some modified gravity parameters have been constrained and thus a case is made that modified gravity can be distinguished from GR by analyzing lensing observables in the strong gravity regime~\cite{2022PhRvD.106f4012K}.

In this work, we have studied gravitational lensing with a simple modified gravity metric. We have considered the SdS metric with $\Psi$ being the modified gravity parameter and thereby obtained the differential time delay between the two lensed images. Note that we have only considered the case of strong lensing where multiple images can be formed. We have found that for the two lensed images, the time difference between these two rays to reach the observer is reduced as the strength of modified gravity increases. We further use this expression to constrain the fraction of dark matter that is made up of primordial black holes using the observed CHIME FRBs. We have considered 636 FRBs and using their reported excess DM values based on the NE2001 model, we have estimated redshifts of the sources, which we further utilize to obtain their integrated optical depths. Finally, we have obtained the $f_\mathrm{PBH}$ bound and found that for a given mass, this bound gets weaker as the strength of modified gravity increases.

It was previously established that $f_\mathrm{PBH}$ bound gets weaker as the lensing due to intergalactic plasma is taken into account~\cite{2022PhRvD.106d3017L}. This means that the lensing due to modified gravity acts similarly to plasma lensing, introducing decoherence to the signal. If any astronomical survey shows a deficit of plasma in the direction of an FRB, modified gravity can certainly make up for it. This result will be improved by the Hydrogen Intensity and Real-time Analysis eXperiment (HIRAX), which will soon be able to detect a lot more FRBs in the Southern sky~\cite{2022JATIS...8a1019C}. Similarly, other telescopes such as Deep Synoptic Array~(DSA)-2000~\cite{2023MNRAS.521.4024C} and Bustling Universe Radio Survey Telescope in Taiwan~(BURSTT)~\cite{2023ApJ...950...53H} are expected to detect gravitationally lensed FRBs to obtain new constraints on $f_\mathrm{PBH}$. A more precise way of modifying the result is by considering a more appropriate metric using the Horndeski or Chameleon gravity following the recent astronomical and cosmological observations. In the future, if the intergalactic plasma contribution can be measured and if any lensing in FRBs is observed, these will help in effectively constraining the various modified gravity models and parameters.


\acknowledgments

We thank the anonymous reviewer for their useful suggestions to improve the quality of the manuscript. S.K. would like to thank Mawson W. Sammons of Curtin University, Zarif Kader of McGill University, Calvin Leung of the Massachusetts Institute of Technology, and Ranjan Laha of the Indian Institute of Science for valuable discussions on FRB lensing. We gratefully acknowledge support from the University of Cape Town Vice Chancellor’s Future Leaders 2030 Awards programme which has generously funded this research and support from the South African Research Chairs Initiative of the Department of Science and Technology and the National Research Foundation. A.W. would like to acknowledge support from the ICTP through the Associates Programme and from the Simons Foundation through grant number 284558FY19. Computations were performed using facilities provided by the University of Cape Town’s ICTS High Performance Computing team: \href{https://ucthpc.uct.ac.za/}{hpc.uct.ac.za}.

\paragraph{Note added.} The data underlying this paper will be shared on a reasonable request to the corresponding author.


\bibliographystyle{JHEP}
\bibliography{bibliography}

\end{document}